\documentclass[a4paper,pdf,11pt]{article}
\usepackage{pos}

\title{NNLO Matrix-Element Corrections in VINCIA}

\author*[a,b]{Peter Skands}
\author[c,d]{Christian Preuss}

\affiliation[a]{School of Physics and Astronomy, Monash University,
 Wellington Road, Clayton, VIC-3800, Australia}

\affiliation[b]{Rudolf Peierls Centre for Theoretical Physics, 
  University of Oxford, Parks Road, Oxford, OX1 3PU, UK}

\affiliation[c]{ Department of Physics, University of Wuppertal, 42119 Wuppertal, Germany}

\affiliation[d]{Institute for Theoretical Physics, ETH, CH-8093 Zürich, Switzerland}

\emailAdd{peter.skands@cern.ch}

\abstract{
We report on a new formalism for parton showers whose fixed-order
expansion can be corrected through next-to-next-to-leading order
(NNLO) in QCD. It is the first such formalism we are aware of that has
no dependence on any auxiliary scales or external resummations and
which is fully differential in all of the relevant phase
spaces. Since the shower acts as the phase-space generator, the
dominant singularity structures are encoded by construction and the
method can generate unweighted events with very high
efficiency without any significant initialisation time. We argue that
the the method should be capable of achieving (at least) NNLO+NNDL accuracy for
the shower evolution variable and use hadronic $Z$ decays as a
specific example.  
}

\FullConference{16th International Symposium on Radiative Corrections: Applications of Quantum Field Theory to Phenomenology (
RADCOR2023)\\
28th May - 2nd June, 2023\\
Crieff, Scotland, UK\\}

\begin{document}
\maketitle

\section{Introduction}
The presence of infrared (IR)
poles in amplitudes with partons that can become soft and/or
collinear complicates making precise predictions in
theories with massless gauge bosons (such as QED and QCD).
Although the resulting IR singularities can be treated
consistently and cancel order by order in the relevant gauge
coupling(s), they leave a legacy in physical observables in
the form of logarithms of scale ratios. If
significant scale  hierarchies are present in the process or 
observables at hand, these logarithms counteract the naive
coupling-power suppression of higher-order terms. This
reduces the effective accuracy of fixed-order calculations for
multi-scale problems.

This is  a concern for ongoing experimental and phenomenological 
studies, e.g.\ at the LHC, where ever-more complex final states are being
targeted --- and accurately measured --- with multiple resolved
objects each of which defines an intrinsic scale, and/or for 
observables sensitive to  substructure. It also applies to differential
observables that cover a wide range of scales over their   
domain(s), which are often well described by fixed-order 
perturbation theory in hard tails while log-enhanced terms affect
the bulk/peak of the differential distributions.

To give a schematic example, an NNLO QCD calculation of a cross section
with a jet veto would include the following terms:
\begin{equation}
    \overbrace{F_0}^{\textcolor{blue}{\mathrm{LO}}}
  ~+~\overbrace{\alpha_s(L^2 + L + F_1)}^{\textcolor{blue}{\mathrm{NLO}}} ~+~ \overbrace{\alpha_s^2 (L^4 + L^3 + L^2 + L + F_2)}^{\textcolor{blue}{\mathrm{NNLO}}}~,
\label{eq:veto}
\end{equation}
where $\alpha_s$ is the QCD coupling constant, $F_i$ denote non-log
terms at each order and $L^m$ in this example represents terms
proportional to powers of logs of the jet-veto scale to a scale
characteristic of the Born-level hard process. If the scales are such that
$\alpha L^2 \sim 1$ then all terms $\alpha_s^n L^{2n}$ would be of
order unity, invalidating any fixed-order truncation of the series. For less extreme hierarchies, the consequence is a reduction of
the effective relative accuracy of the truncation.

At face value, fixed-order calculations are therefore always
most accurate for single-scale problems, while their effective accuracy for
processes/observables with scale hierarchies is reduced. 

The applicability of perturbation
theory can be extended to multi-scale problems by \emph{resumming} the
log-enhanced terms to all 
orders, now using a logarithmic order counting in which a rate like that in 
eq.~(\ref{eq:veto}) is (re)expressed, here shown schematically up to
NNLO+N4DL accuracy: 
\begin{eqnarray}
  \left( \overbrace{F_0}^{\textcolor{blue}{\mathrm{LO}}}
  ~+~ \overbrace{\alpha_s F_1}^{\textcolor{blue}{\mathrm{NLO}}}
  ~+~ \overbrace{\alpha_s^2
    F_2}^{\textcolor{blue}{\mathrm{NNLO}}}\right)
  & \times & \!\exp\Bigg( ~ \overbrace{-\alpha_s
    L^2}^{\textcolor{red}{\mathrm{DL}}} ~ ~\overbrace{ -\alpha_sL ~-
    \alpha_s^2 L^3 }^{ \textcolor{red}{\mathrm{NDL}}} ~ ~\overbrace{ -
    \alpha_s^2 L^2 ~- \alpha_s^3 L^4 }^{\textcolor{red}{\mathrm{NNDL}}}
    \label{eq:resum} \\[-4mm]
 && \qquad ~\underbrace{ - \alpha_s^2 L ~- \alpha_s^3 L^3 ~- \alpha_s^4 L^5
  }_{\textcolor{red}{\mathrm{N3DL}}}
  ~~~\underbrace{ - \alpha_s^3 L^2 ~- \alpha_s^4 L^4 ~- \alpha_s^5 L^6
  }_{\textcolor{red}{\mathrm{N4DL}}}
  \Bigg)~,\nonumber
\end{eqnarray}
where the ``double-log'' (DL) counting in the exponent here is
intended to emphasise that we focus on towers of logs that dominate
in kinematical regions in which $\alpha_s L^2 \sim 1$ (as
distinct from the widely used N$^n$LL counting which is
based on $\alpha_s L \sim 1$).
The fixed-order cofficient $F_1$ is needed both for NLO matching   
and also for NNDL accuracy, and the coefficient $F_2$ is required for
matching to NNLO and for N4DL accuracy. In shower
parlance, exponentials such as the one in eq.~(\ref{eq:resum}) are
called Sudakov factors; we call them that below.

Several general resummation methods exist, which operate at different 
levels of inclusiveness; here we focus only on the most 
exclusive one, parton showers. The requirement of full exclusivity
comes at a cost: this is generally the hardest method to reach
high formal accuracy with. One might ask why, then, pursue this method,
when complementary, more inclusive methods, are available, which can
reach better accuracy than showers do? There are at least four strong
reasons for this:
\begin{enumerate}
\item {\bf Universality.} Given a starting scale, a parton-shower algorithm
  can be applied to \emph{any} parton configuration. This means that,
  once a shower algorithm has been defined and encoded in a
  Monte-Carlo implementation, it can be applied to almost any
  conceivable
  process type, within and beyond the SM, with little or no additional
   manpower. This is the basis of the near-ubiquitous
  applicability of general-purpose shower Monte Carlos (GPMCs) in HEP. 
\item {\bf Efficiency.} Since shower algorithms are based directly on the
  dominant singularity structures of radiative 
  corrections, they are highly efficient in producing unweighted events in the 
  $(\mathrm{Born}+n)$-parton phase spaces. This happens by
  construction, without significant inititalization time. This
  property also underpins 
  so-called forward-branching phase-space
  generators~\cite{Draggiotis:2000gm,Figy:2018imt}.
\item {\bf Fully differential final states.} While 
  inclusive resummation methods typically
  require a separate dedicated calculation for each specific
  observable, shower algorithms produce fully-differential exclusive
  final states, on which \emph{any} observable can be evaluated. Thus,
  one calculation can suffice to make predictions for any number of
  observables. 
\item {\bf The IR cutoff} of the shower algorithm, combined with the fact that
  all-orders corrections have been included above it, makes it
  possible to interface the
  perturbative calculation consistently with explicit and detailed
  dynamical models of hadronization, such as string or cluster
  fragmentation. This in turn also enables embedding the calculation
  within a more  complete modelling framework, including detailed
  simulations of experimental fiducial and efficiency effects, and
  making the calculation accessible to the full suite of 
  collider-physics phenomenology study tools, again a main reason for
  the wide use of GPMCs in HEP. 
\end{enumerate}

These properties, together with the increasing phenomenological
relevance of multi-scale problems in general, make it interesting to
embed fixed-order calculations systematically
within shower calculations, in a  general and efficient way.

Up to NLO accuracy, this is relatively
straightforward~\cite{Bengtsson:1986gz,Norrbin:2000uu,Nason:2004rx,Frixione:2007vw,Alioli:2010xd,Frixione:2002ik,Alwall:2014hca,Frederix:2023hom}. Beyond
NLO, however, there is \emph{ab initio} a problem. At best, current
parton showers achieve NLL resummation
accuracy~\cite{Catani:1990rr,Dasgupta:2018nvj,Dasgupta:2020fwr,Herren:2022jej}. 
Comparing eqs.~(\ref{eq:veto}) and (\ref{eq:resum}), we see that the
$\alpha_s^2 L^2$ and  $\alpha^2_s L$ pieces in eq.~(\ref{eq:veto})
are associaed with NNDL and N3DL terms in eq.~(\ref{eq:resum}),
respectively; these are categorised as NLL and NNLL respectively if
one employs ``Caesar-style'' log counting~\cite{Hamilton:2020rcu}. 
This makes it impossible to write down matching
equations in which the log-enhanced terms are all on the shower
side. To circumvent this issue, previous NNLO matching 
approaches~\cite{Hamilton:2013fea,Alioli:2013hqa,Alioli:2015toa,Monni:2019whf} 
have utilised analytically 
calculated Sudakov factors to supplement the parton-shower ones. 
This is probably the best one can do with current showers but does
have the drawback that the accuracy in shower-dominated phase-space 
regions is not improved. There is also the need
to calculate separate analytical Sudakov factors. And
subtleties associated with the fact that they (and their
resummation variables) are not completely identical to their
equivalents on the shower side, though this difference can be made at
least formally subleading by making suitable 
resummation-variable and scale choices. 

Returning to eqs.~(\ref{eq:veto}) and (\ref{eq:resum}), a fully
self-contained embedding of an NNLO calculation in a parton-shower
framework would appear to require an N3DL accurate shower
algorithm (and N3LO calculations, which are also beginning to emerge
on the phenomenological scene, would then require N6DL showers). This is
not realistic to shoot for, and is also not strictly
necessary.

Instead, we aim for a consistent
shower that exponentiates the full ${\cal O}(\alpha_s^2)$ pole
structure of the NNLO fixed-order matrix elements. This is sufficient
to enable a fully differential matching, where all poles that appear
on the fixed-order side also appear on the shower side.
If the ${\cal O}(\alpha_s^3)$ soft anomalous dimension is also
included, we argue that the shower Sudakov factor contains
all terms required for NNDL accuracy on the shower evolution
variable. By construction, the method also
exponentiates the N3DL $\alpha_s^2 L$ term, but the other N3DL
coefficients are not included. We note that for modest scale
hierarchies, characterised by $\alpha_s L^2 < 1$,
the relative importance of the $\alpha_s^3 L$ and $\alpha_s^2 L$
cofficients swap places, hence in such regions we would still expect
our partial N3DL resummation to represent a systematic improvement
over NNDL.  
 
Below, we describe the ingredients that are needed to accomplish
this, based on
refs.~\cite{Hartgring:2013jma,Li:2016yez,Brooks:2020upa,Campbell:2021svd}.  

\section{Phase-Space Generation and NNLO Matching}

In a conventional fixed-order calculation, each of the
$(\mathrm{Born}+m)$-parton phase spaces are generated separately.
In a shower-style algorithm, instead all events start out
as Born-level events, and all higher multiplicities are produced by
the shower branching process. The unitarity of the 
shower generates a Sudakov-weighting of exclusive cross
sections, which at each higher multiplicity comes multiplied by the
kernel(s) of the relevant branchings. If the shower algorithm is
sufficiently tractable, these weights can be
expanded and matched to any given 
fixed order~\cite{Giele:2011cb}. This has been worked out for
final-state antenna showers at both tree
level~\cite{Giele:2011cb,Lopez-Villarejo:2011pwr,Larkoski:2013yi}
and at one loop~\cite{Hartgring:2013jma}. 

Here, we focus on a MEC/POWHEG-style multiplicative matching
procedure. For this to work, it is obviously 
necessary that the shower algorithm is able to populate all of the relevant
phase spaces, with no ``dead zones''. This is not true
of conventional strongly-ordered parton showers\footnote{This can
in principle be circumvented by modifying the shower ordering
variable (e.g., virtuality-ordering can trivially be seen to cover
all of phase space), but at least for LL branching kernels we are 
discouraged from doing so, for the reasons elaborated on
in ref.~\cite{Giele:2011cb}, and it also appears to lead to the wrong
resummation 
structure~\cite{Hartgring:2013jma,Li:2016yez}. Another option
was ``smooth ordering''~\cite{Giele:2011cb}, but again we believe 
this would lead to an undesirable resummation structure.}. E.g., a
$p_\perp$-ordering condition will typically cut out part of the
$(\mathrm{Born}+2)$-parton phase space~\cite{Giele:2011cb}. 

A path to a robust approach can be found by analysing the
propagator structure of the amplitudes that contribute to the regions
that are cut out by strong ordering. These phase-space points are
characterised by having no strong hierarchy in the propagator
virtualities. Intuitively, they should therefore be thought of not as
resulting from iterated (ordered) $n\to n+1$ splittings, but as direct 
(single-scale) $n \to n+2$ splittings. They are also associated with
qualitatively different terms in both the fixed-order and logarithmic
expansions than the points in the ordered region are; the unordered
region only borders on double-unresolved limits of the fixed-order
matrix elements, and hence integrals over it should also only
contribute to $\alpha^2_s L^2$ (NNDL) and $\alpha_s L$ (N3DL)
coefficients. This is consistent with conventional showers being able
to reach up to NDL accuracy without addressing this region, but we
suspect it would not be possible to reach accuracy higher than NDL 
without some form of dedicated treatment of the
unordered/double-unresolved region of phase space.

Our solution~\cite{Li:2016yez} is to add new ``direct'' $2\to 4$
branchings, based on a Sudakov-style 6D phase-space sampler with an 
$\alpha_s^2/p_\perp^4$ kernel. In a sector-shower 
context~\cite{Lopez-Villarejo:2011pwr,Brooks:2020upa}, we divide the
$(\mathrm{Born}+2)$-parton 
phase-space cleanly into an unordered sector to be populated 
by the direct $2\to 4$ sampler, and an ordered sector populated by 
 iterated $2\to 3$ ones. (In a global shower, one would instead
sum over  $2\to 3$ and $2\to 4$ contributions, when formulating the
matching conditions.) 
The specific criterion we use to decide which sector we are in is the
following: in an $m$-parton configuration, find the smallest
(colour-ordered) 3-parton $p_\perp$ resolution scale. Tentatively
perform that clustering, using antenna kinematics. Now again find the
smallest 3-parton $p_\perp$ resolution scale, denoted
$\hat{p}_\perp$. If $\hat{p}_\perp > p_\perp$, the $(m \to
m-1)$-parton clustering is ordered; otherwise it is unordered.

\paragraph{Unordered Part:} to realise the direct $2\to 4$ sampler, we
make use of the 
iterated (exact) $2\to 3$ antenna phase-space factorisation, and
define a \emph{trial} $2\to 4$ Sudakov factor as follows: 
\begin{equation}
  -\ln \hat{\Delta}_{2\to 4}(p_{\perp 0}^2, p_\perp^2) =
  \int_0^{p_{\perp 0}^2} \mathrm{d} p_{\perp 1}^2
  \int_{p_\perp^2}^{p_{\perp 0}^2} 
  \mathrm{d} p_{\perp 2}^2 \, \overbrace{\Theta( p_{\perp 2}^2 -
    p_{\perp 1}^2)}^{\mathrm{\textcolor{magenta}{Unordered:~p_{\perp1}
    < p_{\perp 2}}}} \,
  \int \!
  \mathrm{d}y_1 \mathrm{d}y_2 \, \hat{a}_{2\to 4}~,
\end{equation}
where a simple choice for the trial function $\hat{a}_{2\to 4}$ is
proportional to $C_A^2 \,\alpha_s^2 (p_{\perp 2}^2) / p_{\perp 2}^4$.
We then exploit the definition of the unordered region to swap the order of the integrations,
\begin{eqnarray}
  \implies ~~\,-\ln \hat{\Delta}_{2\to 4}(p_{\perp 0}^2, p_\perp^2) & = &
  \int_{p_\perp^2}^{p_{\perp 0}^2} \mathrm{d}p_{\perp 2}^2
  \underbrace{\int_0^{p_{\perp 2}^2} \mathrm{d} p_{\perp
      1}^2}_{\mathrm{\textcolor{magenta}{Unordered: p_{\perp1}
    < p_{\perp 2}}}} \int \!
    \mathrm{d}y_1 \mathrm{d}y_2 \, \hat{a}_{2\to 4} ~.
\end{eqnarray}
Details of the trial-generation procedure are given in
ref.~\cite{Li:2016yez}. Unweighted ME-corrected
events are generated by accepting trial branchings with a tree-level
second-order MEC ratio, 
\begin{equation}
  P^{\mathrm{MEC}}_{2\to 4} ~=~ 
  \frac{|M_{\mathrm{Born+2}}^{(0)}|^2}{\hat{a}_{2\to 4}
    |M_{\mathrm{Born}}^{(0)}|^2}~,
  \label{eq:mec2to4}
\end{equation}
where subscripts denote multiplicities and the superscript indicates
relative loop order. For hadronic $Z$ decay, the physical $2\to 4$ Sudakov
factor generated by the matched shower is then: 
\begin{equation}
  -\ln \Delta_{2\to 4}(m_Z^2, p_\perp^2) ~ =~ 
  \int_{p_\perp^2}^{m_Z^2} \mathrm{d}p_{\perp 2}^2
  \int_0^{p_{\perp 2}^2} \mathrm{d} p_{\perp
      1}^2 \int \!
    \mathrm{d}y_1 \mathrm{d}y_2 \,
    \frac{|M^{(0)}_\mathrm{q\bar{q}gg}|^2}{|M^{(0)}_{q\bar{q}}|^2} 
    ~.\label{eq:sud2to4}
\end{equation}
After a $2\to 4$ trial branching is accepted, the pure shower
evolution can simply be continued, starting from the $p_\perp$ scale
of the accepted $2\to 4$ branching. 

\paragraph{Ordered Part:} in the ordered part of the nested phase
spaces, the first $2\to 3$ branching receives a standard first-order
(tree-level) MEC, augmented by a second-order (one-loop) correction, 
\begin{equation}
  P^{\mathrm{MEC}}_{2\to 3} ~=~
  \overbrace{\frac{|M_{\mathrm{Born+1}}^{(0)}|^2}{\hat{a}_{2\to 3}
    |M_{\mathrm{Born}}^{(0)}|^2}}^{\textcolor{orange}{\mathrm{Tree-Level~}2\to
      3~\mathrm{MEC}}}  \overbrace{\left(1 ~+~
    \tilde{w}^\mathrm{NLO}_{\mathrm{Born}+1}
    ~ - ~ \tilde{w}^\mathrm{NLO}_{\mathrm{Born}} 
    ~ - ~ \tilde{w}^\mathrm{Sudakov}_{2\to 3}
    ~ - ~
    \frac{\alpha_s}{2\pi}\frac{\beta_0}{2}\ln\frac{\mu_\mathrm{FO}^2}{\mu_\mathrm{PS}^2}
    \right)}^{\textcolor{orange}{\mathrm{One-Loop~Corrections}}}~.
  \label{eq:w3NLO}
\end{equation}
The one-loop corrections are defined so that the second-order
shower expansion will match the NNLO real-virtual
coefficient~\cite{Campbell:2021svd}. 
The fixed-order weights $\tilde{w}^{\mathrm{NLO}}_{\mathrm{Born+1}}$
and $\tilde{w}^{\mathrm{NLO}}_{\mathrm{Born}}$ are each IR finite, 
\begin{equation}
|M_{\mathrm{Born}+m}^{(0)}|^2\,\tilde{w}^\mathrm{NLO}_{\mathrm{Born}+m}  ~=~ 
2\mathrm{Re}\big[M_{\mathrm{Born}+m}^{(1)}M_{\mathrm{Born}+m}^{(0)*}\big]
  \,+\, \int_0^{\mathrm{p_{\perp m}^2}} \mathrm{d}\Phi_{+1}
  \,|M_{\mathrm{Born}+m+1}^{(0)}|^2 ~,
\end{equation}
and $\tilde{w}^\mathrm{Sudakov}_{2\to 3}$ is the first-order expansion of 
the $2\to 3$ shower Sudakov weight~\cite{Hartgring:2013jma},
\begin{equation}
\tilde{w}^\mathrm{Sudakov}_{2\to 3} ~=~
~-~ \int_{p_\perp^2}^{p_{\perp 0}^2} \mathrm{d}\Phi_{+1}
\frac{|M_{\mathrm{Born}+1}^{(0)}|^2}{|M_{\mathrm{Born}}^{(0)}|^2}~.
\end{equation}
The last term in eq.~(\ref{eq:w3NLO}) matches the parton-shower
and fixed-order renormalisation-scale choices. The canonical
choice for coherent showers is $\mu_R \propto p_\perp$ augmented by the
so-called ``CMW factor'', $\kappa_\mathrm{CMW}$~\cite{Catani:1990rr}, which
absorbs the 2-loop cusp anomalous dimension, 
\begin{equation}
  \mu^2_\mathrm{PS} = \kappa_\mathrm{CMW}^2 p_\perp^2 ~~~,~~~ \kappa^2 = \exp(
  K/\beta_0 ) ~~~,~~~ K = \frac{67 C_A}{18} - \frac{\pi^2}{6}
  - \frac{10 n_F T_R}{9}  ~~~,~~~\beta_0 = \frac{11 C_A - 4 T_R n_F}{3}~.  
\end{equation}

Putting it all together, the $2\to 3$ Sudakov factor for
hadronic $Z$ decay becomes:
\begin{eqnarray}\label{eq:sud2to3}
  -\ln\Delta_{2\to 3}(m_Z^2,p_\perp^2) & = & \int_{p_\perp^2}^{m_Z^2}
  \mathrm{d}p_{\perp 1}^2 \mathrm{d} y_1 ~\Bigg(~ 
    \frac{|M_{q\bar{q}g}^{(0)}|^2}{|M_{q\bar{q}}^{(0)}|^2}\,\Big[\,1
    \,- \overbrace{\frac{\alpha_s}{\pi}}^{\textcolor{brown}{\tilde{w}^\mathrm{NLO}_\mathrm{Born}}} 
    +\, \overbrace{\int_{p_\perp^2}^{m_Z^2} \mathrm{d}p_\perp'^2 \mathrm{d}y'
\frac{|M_{q\bar{q}g'}^{(0)}|^2}{|M_{q\bar{q}}^{(0)}|^2}}^{\textcolor{brown}{\tilde{w}^\mathrm{Sudakov}_{2\to
          3}}}  \\[1.5mm]
& &   \hspace*{-0.2cm}-\,\frac{\alpha_s}{2\pi}\frac{\beta_0}{2}\ln\frac{m_Z^2}{\kappa^2
      p_{\perp 1}^2}\,\Big]+\, \underbrace{\frac{2\mathrm{Re}\big[M_{q\bar{q}g}^{(1)}M_{q\bar{q}g}^{(0)*}\big]}{|M_{q\bar{q}}^{(0)}|^2}
  \,+\, \int_0^{\mathrm{p_{\perp 1}^2}} \mathrm{d}p_{\perp 2}^2 \mathrm{d}y_{2}
  \,\frac{|M_{q\bar{q}gg}^{(0)}|^2}{|M_{q\bar{q}}^{(0)}|^2}}_{\textcolor{brown}{\tilde{w}^\mathrm{NLO}_\mathrm{Born+1}}}\,\Bigg)\,.\nonumber
\end{eqnarray}

The combined Sudakov factor for a
jet veto at the scale $p_\perp$ is found by multiplying the two
Sudakov factors, eqs.~(\ref{eq:sud2to4}) and
eq.~(\ref{eq:sud2to3}). We see that something quite beautiful
happens; the integral over ordered 4-parton phase-space points in the
last term of eq.~(\ref{eq:sud2to3}) combines with that over
unordered 4-parton points in eq.~(\ref{eq:sud2to4}) to yield a seamless
integral over the full 4-parton phase space. 

Before analysing the structure of the combined Sudakov factor in more
detail, two final aspects must be clarified to define the full NNLO
matching. The first is that, after a $2\to 3$ trial branching is
accepted, the shower evolution continues, starting from the $p_\perp$ scale
of the accepted $2\to 3$ branching. A tree-level 4-parton ME
correction factor is then applied to the next branching, analogous to
that used in the unordered region, eq.~(\ref{eq:mec2to4}), but here
for ordered histories:
\begin{equation}
  P^{\mathrm{MEC}}_{3\to 4} ~=~ 
  \frac{|M_{\mathrm{Born+2}}^{(0)}|^2}{\hat{a}_{3\to 4} |M_{\mathrm{Born+1}}^{(0)}|^2}~.
\end{equation}
Thus all 4-parton points are corrected to the NNLO matrix element,
irrespective of whether they are reached by the direct
$2\to 4$ generator, or by the iterated $2\to 3$ generator. 

The second aspect of achieving the NNLO matching is that, since all
events start out as Born-level events, the Born-level phase-space
weight is augmented by a differential NNLO ``K-factor'', which
enforces the NNLO normalisation of the total cross section and differential
distributions~\cite{Campbell:2021svd}. 

\section{Argument for NNDL Accuracy in the Shower $\mathbf{p_\perp}$ Evolution Variable}

Let us be a bit more definite about what exactly the combined Sudakov
factor corresponds to. Specifically, consider a jet clustering
algorithm that corresponds to the inverse of 
the sector-shower branching algorithm. Since we use dipole-antenna
kinematics and ARIADNE $p_\perp$~\cite{Gustafson:1987rq} as our
sector-resolution variable, this is known as the ARCLUS
algorithm~\cite{Lonnblad:1992qd}, suitably extended to 
incorporate inverses of our new direct $2\to 4$ branchings. We call
this ARCLUS 2.

For a global shower, this jet algorithm would have to be defined in a
stochastic way, to allow for the multiple histories that can
contribute to each phase-space point. But since a sector shower is
bijective the corresponding inverse algorithm is in our case a
conventional deterministic jet clustering algorithm, 
producing a unique clustering sequence for each event.

The rate of events that will pass an ARCLUS-2 jet veto  at a scale
$p_{\perp}$ is: 
\begin{equation}
  k^\mathrm{NNLO}\, |M_\mathrm{Born}^{(0)}|^2\, \Delta_{2\to 3}(m^2_Z,
  p^2_{\perp})\,\Delta_{2\to
    4}(m^2_Z,p^2_{\perp})~.
  \label{eq:vinciaVeto}
\end{equation}
We shall assume that the NNLO matching ensures that $k^\mathrm{NNLO}$
matches the coefficients $F_i$ in eq.~(\ref{eq:resum}).

Before considering the log terms in the Sudakov
factors, we first ask whether further shower evolution could in
principle lead to violations of the jet veto, e.g., via recoil
effects from subsequent branchings. If so, that would invalidate
eq.~(\ref{eq:vinciaVeto}). In a global shower setup, this question is 
nontrivial since at least some of the shower histories would
involve scales higher than the veto scale, and because recoils that
increase the resolution scale are not explicitly forbidden. In a
sector shower setup, however, neither of these complications are
present, hence the above equation is exact. 

Assuming the order-$\alpha_s$ log terms to be guaranteed by the
integral over the tree-level matrix-element ratio,
$|M_{q\bar{q}g}|^2/M_{q\bar{q}}|^2$ and the remaining $\alpha_s^2 L^3$
NDL coefficient via the CMW factor, the question of NNDL accuracy on
eq.~(\ref{eq:vinciaVeto}) boils down to whether the remaining terms
in the combined shower Sudakov produce the correct $\alpha_s^2 L^2$
coefficient in eq.~(\ref{eq:resum}). We then rely on extending
the CMW prescription to match the 3-loop cusp anomalous dimension to
get the $\alpha_s^3 L^4$ piece.

Terms proportional to $\alpha_s^2 L^2$ arise in quite a few places in
eqs.~(\ref{eq:sud2to3}) and (\ref{eq:sud2to4}). Many of these are
analytically tractable, e.g.\ using the expressions in
\cite{Gehrmann-DeRidder:2005btv,Hartgring:2013jma}; the most
challenging are the ones from the 4-parton phase space. We have not
completed a full analysis of this structure yet and hence are not in a
position to \emph{prove} NNDL accuracy. However, since all of the
relevant ME poles are clearly exponentiated in eqs.~(\ref{eq:sud2to3})
and (\ref{eq:sud2to4}), with the matching to fixed order eliminating
double-counting of non-singular coefficients (like $\alpha_s/\pi$), we
believe there is good reason to expect that the method we have
proposed is capable of achieving (at least) NNDL accuracy.

For clarity and completeness, we emphasise that we are only making
this statement about an observable that corresponds to the
shower-evolution variable itself. We also note that we have here
neglected subtleties that arise at subleading colour. 

\subsubsection*{Acknowledgments}
We are grateful to L.~Scyboz and to B.~El-Menoufi for helpful comments on
the draft. This work was supported by the Australian Research Council
Discovery Project DP220103512 ``Tackling the Computational Bottleneck
in Precision Particle Physics''. 

\bibliographystyle{JHEP}
\bibliography{skands}

\providecommand{\href}[2]{#2}\begingroup\raggedright\begin{thebibliography}{10}

\bibitem{Draggiotis:2000gm}
P.~D. Draggiotis, A.~van Hameren and R.~Kleiss, \emph{{SARGE: An Algorithm for
  generating QCD antennas}},
  \href{https://doi.org/10.1016/S0370-2693(00)00532-3}{\emph{Phys. Lett. B}
  {\bfseries 483} (2000) 124}
  [\href{https://arxiv.org/abs/hep-ph/0004047}{{\ttfamily hep-ph/0004047}}].

\bibitem{Figy:2018imt}
T.~M. Figy and W.~T. Giele, \emph{{A Forward Branching Phase Space Generator
  for Hadron colliders}},
  \href{https://doi.org/10.1007/JHEP10(2018)203}{\emph{JHEP} {\bfseries 10}
  (2018) 203} [\href{https://arxiv.org/abs/1806.09678}{{\ttfamily
  1806.09678}}].

\bibitem{Bengtsson:1986gz}
M.~Bengtsson, T.~Sj{\"o}strand and M.~van Zijl, \emph{{Initial State Radiation
  Effects on $W$ and Jet Production}},
  \href{https://doi.org/10.1007/BF01441353}{\emph{Z. Phys. C} {\bfseries 32}
  (1986) 67}.

\bibitem{Norrbin:2000uu}
E.~Norrbin and T.~Sj{\"o}strand, \emph{{QCD radiation off heavy particles}},
  \href{https://doi.org/10.1016/S0550-3213(01)00099-2}{\emph{Nucl. Phys. B}
  {\bfseries 603} (2001) 297}
  [\href{https://arxiv.org/abs/hep-ph/0010012}{{\ttfamily hep-ph/0010012}}].

\bibitem{Nason:2004rx}
P.~Nason, \emph{{A New method for combining NLO QCD with shower Monte Carlo
  algorithms}},
  \href{https://doi.org/10.1088/1126-6708/2004/11/040}{\emph{JHEP} {\bfseries
  11} (2004) 040} [\href{https://arxiv.org/abs/hep-ph/0409146}{{\ttfamily
  hep-ph/0409146}}].

\bibitem{Frixione:2007vw}
S.~Frixione, P.~Nason and C.~Oleari, \emph{{Matching NLO QCD computations with
  Parton Shower simulations: the POWHEG method}},
  \href{https://doi.org/10.1088/1126-6708/2007/11/070}{\emph{JHEP} {\bfseries
  11} (2007) 070} [\href{https://arxiv.org/abs/0709.2092}{{\ttfamily
  0709.2092}}].

\bibitem{Alioli:2010xd}
S.~Alioli et~al., \emph{{A general framework for implementing NLO calculations
  in shower Monte Carlo programs: the POWHEG BOX}},
  \href{https://doi.org/10.1007/JHEP06(2010)043}{\emph{JHEP} {\bfseries 06}
  (2010) 043} [\href{https://arxiv.org/abs/1002.2581}{{\ttfamily 1002.2581}}].

\bibitem{Frixione:2002ik}
S.~Frixione and B.~R. Webber, \emph{{Matching NLO QCD computations and parton
  shower simulations}},
  \href{https://doi.org/10.1088/1126-6708/2002/06/029}{\emph{JHEP} {\bfseries
  06} (2002) 029} [\href{https://arxiv.org/abs/hep-ph/0204244}{{\ttfamily
  hep-ph/0204244}}].

\bibitem{Alwall:2014hca}
J.~Alwall et~al., \emph{{The automated computation of tree-level and NLO
  differential cross sections, and their matching to parton shower
  simulations}}, \href{https://doi.org/10.1007/JHEP07(2014)079}{\emph{JHEP}
  {\bfseries 07} (2014) 079} [\href{https://arxiv.org/abs/1405.0301}{{\ttfamily
  1405.0301}}].

\bibitem{Frederix:2023hom}
R.~Frederix and P.~Torrielli, \emph{{A new way of reducing negative weights in
  MC@NLO}},  \href{https://arxiv.org/abs/2310.04160}{{\ttfamily 2310.04160}}.

\bibitem{Catani:1990rr}
S.~Catani, B.~R. Webber and G.~Marchesini, \emph{{QCD coherent branching and
  semiinclusive processes at large x}},
  \href{https://doi.org/10.1016/0550-3213(91)90390-J}{\emph{Nucl. Phys. B}
  {\bfseries 349} (1991) 635}.

\bibitem{Dasgupta:2018nvj}
M.~Dasgupta, F.~A. Dreyer, K.~Hamilton, P.~F. Monni and G.~P. Salam,
  \emph{{Logarithmic accuracy of parton showers: a fixed-order study}},
  \href{https://doi.org/10.1007/JHEP09(2018)033}{\emph{JHEP} {\bfseries 09}
  (2018) 033} [\href{https://arxiv.org/abs/1805.09327}{{\ttfamily
  1805.09327}}].

\bibitem{Dasgupta:2020fwr}
M.~Dasgupta, F.~A. Dreyer, K.~Hamilton, P.~F. Monni, G.~P. Salam and G.~Soyez,
  \emph{{Parton showers beyond leading logarithmic accuracy}},
  \href{https://doi.org/10.1103/PhysRevLett.125.052002}{\emph{Phys. Rev. Lett.}
  {\bfseries 125} (2020) 052002}
  [\href{https://arxiv.org/abs/2002.11114}{{\ttfamily 2002.11114}}].

\bibitem{Herren:2022jej}
F.~Herren, S.~H\"oche, F.~Krauss, D.~Reichelt and M.~Schoenherr, \emph{{A new
  approach to color-coherent parton evolution}},
  \href{https://arxiv.org/abs/2208.06057}{{\ttfamily 2208.06057}}.

\bibitem{Hamilton:2020rcu}
K.~Hamilton, R.~Medves, G.~P. Salam, L.~Scyboz and G.~Soyez, \emph{{Colour and
  logarithmic accuracy in final-state parton showers}},
  \href{https://doi.org/10.1007/JHEP03(2021)041}{\emph{JHEP} {\bfseries 03}
  (2021) 041} [\href{https://arxiv.org/abs/2011.10054}{{\ttfamily
  2011.10054}}].

\bibitem{Hamilton:2013fea}
K.~Hamilton et~al., \emph{{NNLOPS simulation of Higgs boson production}},
  \href{https://doi.org/10.1007/JHEP10(2013)222}{\emph{JHEP} {\bfseries 10}
  (2013) 222} [\href{https://arxiv.org/abs/1309.0017}{{\ttfamily 1309.0017}}].

\bibitem{Alioli:2013hqa}
S.~Alioli et~al., \emph{{Matching Fully Differential NNLO Calculations and
  Parton Showers}}, \href{https://doi.org/10.1007/JHEP06(2014)089}{\emph{JHEP}
  {\bfseries 06} (2014) 089} [\href{https://arxiv.org/abs/1311.0286}{{\ttfamily
  1311.0286}}].

\bibitem{Alioli:2015toa}
S.~Alioli et~al., \emph{{Drell-Yan production at NNLL'+NNLO matched to parton
  showers}}, \href{https://doi.org/10.1103/PhysRevD.92.094020}{\emph{Phys. Rev.
  D} {\bfseries 92} (2015) 094020}
  [\href{https://arxiv.org/abs/1508.01475}{{\ttfamily 1508.01475}}].

\bibitem{Monni:2019whf}
P.~Monni et~al., \emph{{MiNNLO$_{PS}$: a new method to match NNLO QCD to parton
  showers}}, \href{https://doi.org/10.1007/JHEP05(2020)143}{\emph{JHEP}
  {\bfseries 05} (2020) 143}
  [\href{https://arxiv.org/abs/1908.06987}{{\ttfamily 1908.06987}}].

\bibitem{Hartgring:2013jma}
L.~Hartgring, E.~Laenen and P.~Skands, \emph{{Antenna Showers with One-Loop
  Matrix Elements}}, \href{https://doi.org/10.1007/JHEP10(2013)127}{\emph{JHEP}
  {\bfseries 10} (2013) 127} [\href{https://arxiv.org/abs/1303.4974}{{\ttfamily
  1303.4974}}].

\bibitem{Li:2016yez}
H.~T. Li and P.~Skands, \emph{{A framework for second-order parton showers}},
  \href{https://doi.org/10.1016/j.physletb.2017.05.011}{\emph{Phys. Lett. B}
  {\bfseries 771} (2017) 59}
  [\href{https://arxiv.org/abs/1611.00013}{{\ttfamily 1611.00013}}].

\bibitem{Brooks:2020upa}
H.~Brooks, C.~T. Preuss and P.~Skands, \emph{{Sector Showers for Hadron
  Collisions}}, \href{https://doi.org/10.1007/JHEP07(2020)032}{\emph{JHEP}
  {\bfseries 07} (2020) 032}
  [\href{https://arxiv.org/abs/2003.00702}{{\ttfamily 2003.00702}}].

\bibitem{Campbell:2021svd}
J.~M. Campbell et~al., \emph{{Towards NNLO+PS matching with sector showers}},
  \href{https://doi.org/10.1016/j.physletb.2022.137614}{\emph{Phys. Lett. B}
  {\bfseries 836} (2023) 137614}
  [\href{https://arxiv.org/abs/2108.07133}{{\ttfamily 2108.07133}}].

\bibitem{Giele:2011cb}
W.~T. Giele, D.~A. Kosower and P.~Z. Skands, \emph{{Higher-Order Corrections to
  Timelike Jets}},
  \href{https://doi.org/10.1103/PhysRevD.84.054003}{\emph{Phys. Rev. D}
  {\bfseries 84} (2011) 054003}
  [\href{https://arxiv.org/abs/1102.2126}{{\ttfamily 1102.2126}}].

\bibitem{Lopez-Villarejo:2011pwr}
J.~J. Lopez-Villarejo and P.~Z. Skands, \emph{{Efficient Matrix-Element
  Matching with Sector Showers}},
  \href{https://doi.org/10.1007/JHEP11(2011)150}{\emph{JHEP} {\bfseries 11}
  (2011) 150} [\href{https://arxiv.org/abs/1109.3608}{{\ttfamily 1109.3608}}].

\bibitem{Larkoski:2013yi}
A.~J. Larkoski, J.~J. Lopez-Villarejo and P.~Skands, \emph{{Helicity-Dependent
  Showers and Matching with VINCIA}},
  \href{https://doi.org/10.1103/PhysRevD.87.054033}{\emph{Phys. Rev. D}
  {\bfseries 87} (2013) 054033}
  [\href{https://arxiv.org/abs/1301.0933}{{\ttfamily 1301.0933}}].

\bibitem{Gustafson:1987rq}
G.~Gustafson and U.~Pettersson, \emph{{Dipole Formulation of QCD Cascades}},
  \href{https://doi.org/10.1016/0550-3213(88)90441-5}{\emph{Nucl. Phys. B}
  {\bfseries 306} (1988) 746}.

\bibitem{Lonnblad:1992qd}
L.~L{\"o}nnblad, \emph{{ARCLUS: A New jet clustering algorithm inspired by the
  color dipole model}}, \href{https://doi.org/10.1007/BF01557706}{\emph{Z.
  Phys. C} {\bfseries 58} (1993) 471}.

\bibitem{Gehrmann-DeRidder:2005btv}
A.~Gehrmann-De~Ridder, T.~Gehrmann and E.~W.~N. Glover, \emph{{Antenna
  subtraction at NNLO}},
  \href{https://doi.org/10.1088/1126-6708/2005/09/056}{\emph{JHEP} {\bfseries
  09} (2005) 056} [\href{https://arxiv.org/abs/hep-ph/0505111}{{\ttfamily
  hep-ph/0505111}}].

\end{thebibliography}\endgroup

\end{document}